\documentclass[11pt,twocolumn,twoside]{IEEEtran}
\usepackage{amsmath}

\usepackage[margin=0.75in,headheight=0.45in]{geometry}
\usepackage[pdftex]{epsfig}
\usepackage{amsfonts}
\usepackage{amssymb}
\usepackage{fancyhdr}

\usepackage{cite}
\usepackage{float}
\usepackage{subfloat}
\usepackage{subfig}
\usepackage{bm}

\begin{document}
\title{\vspace{0.2in}\sc Machine Learning of Committor Functions for Predicting High Impact Climate Events}
\author{Dario Lucente, Stefan Duffner, Corentin Herbert, Joran Rolland, Freddy Bouchet}

\maketitle
\thispagestyle{fancy}
\begin{abstract}
There is a growing interest in the climate community to improve the prediction of high impact climate events, for instance ENSO (El-Ni\~no--Southern Oscillation) or extreme events, using a combination of model and observation data. In this note we explain that, in a dynamical context, the relevant quantity for predicting a future event is a committor function. We explain the main mathematical properties of this probabilistic concept. We compute and discuss the committor function of the Jin and Timmerman model of El-Ni\~no. Our first conclusion is that one should generically distinguish between states with either intrinsic predictability or intrinsic unpredictability. This predictability concept is markedly different from the deterministic unpredictability arising because of chaotic dynamics and exponential sensibility to initial conditions. The second aim of this work is to compare the inference of a committor function from data, either through a direct approach or through a machine learning approach using neural networks. We discuss the consequences of this study for future applications to more complex data sets.
\end{abstract}

\section{Introduction}
The low frequency modes of variability of the climate system, for instance ENSO \cite{dijkstra2013nonlinear,jin1997equatorial1,jin1997equatorial2,timmermann2003nonlinear,timmermann2002nonlinear,guckenheimer2017predictability,feng2017climate,nooteboom2018using,ludescher2014very}, have a huge impact on nature and human societies, through their local or global signatures. Rare events, such as heat waves, floods, or hurricanes, may also have a huge impact \cite{ragone2018computation,field2012managing,aghakouchak2012extremes,herring2014explaining,coumou2012decade}. Predicting the occurrence of such events is thus a major challenge  \cite{ragone2018computation,field2012managing}. Because the dynamics of the climate system is chaotic, one usually distinguishes between time scales much shorter than a Lyapunov time\footnote{The Lyapunov time is the time required for having a separation between two trajectories which start from very close initial conditions, for more details see \cite{castiglione2008chaos}.} for which a deterministic weather forecast is relevant, and time scales much longer than a mixing times beyond which any deterministic forecast is irrelevant and only climate averaged or probabilistic quantities can be predicted. However, for most applications cited above, the largest interest is for intermediate time scales for which some information, more precise than the climate averages, might be predicted, but for which a deterministic forecast is not relevant. We call this range of time scales {\it the predictability margin}. As far as applications are concerned, many cases of “medium-range forecast” are within the “predictability margin” range. As a paradigmatic example, we study in this work the probability that El-Ni\~no might occur following year. Another example could be: What is the probability of a heat wave of a given amplitude to happen next summer, given the state of the atmosphere, ocean, and soil moisture, in Spring?  For such questions, the system really behaves in a stochastic way as can be seen from the top panel of figure (\ref{ComparisonModelNinoIndex}), which shows the behavior of El-Ni\~{n}o3 anomaly. 

We stress in this work that the prediction problem at the predictability margin is of a probabilistic nature. Indeed, such time scales might typically be of the order of the Lyapunov time scale or larger, where errors on the initial condition and model errors limit our ability to compute deterministically the evolution. However, we stress that the Lyapunov time scale, a global quantity, is clearly not the relevant dynamical quantity for this predictability problem. By contrast, at the predictability margin, the predictability clearly depends on the current state of the system. What is then the relevant mathematical concept? The first aim of this work is to introduce in the field of climate the notion of the committor function \cite{vanden2006transition,weinan2005transition}. A committor function is the probability that an event will occur or not in the future, as a function of the current state of the system. For the El-Ni\~{n}o case, this committor function will be the probability that an observable $\mathcal{O}$ of the system reaches a given threshold within a time $T$ \cite{lestang2018computing}. The definition and mathematical properties of committor functions are introduced in section III.

The first result of this work is to demonstrate, using the committor function, that a predictability margin exists for El-Ni\~{n}o. This demonstration is performed within the Jin and Timmermann model, a low dimensional model proposed to explain the decadal amplitude changes of El-Ni\~no \cite{timmermann2003nonlinear,timmermann2002nonlinear} (section II). From the computed committor function for the Jin and Timmerman model (section III), we obtain the second main result of this work. This result is the characterisation of regions of the phase space with qualitatively different predictability properties. For example for the intermediate stochasticity regime at the predictability margin, we delineate 4 regions, see (Fig. 4b) that will be explain in section \ref{CommittorFunctions}. First, two regions of perfect predictability, where the event will occur with probability $0$ or $1$, respectively. Second, regions with good predictability properties where a value of the probability $0<q<1$ can clearly be predicted with very mild dependance with respect to initial condition. We call this area the {\it probabilistically predictable region}. Third, regions which are unpredictable in practice, because the strong dependance with respect to the initial condition prevent  any practical prediction, either deterministic or probabilistic.  The existence of such features, and especially the new and most interesting {\it probabilistically predictable region}, should be generic for most prediction problems in climate dynamics.

We will explain that committor functions solve Dirichlet problems. However such partial differential equations are extremely difficult to solve especially for high-dimensional systems. Could we compute it directly from data? There is currently a growing interest to estimate relevant dynamical quantities directly from available data, for instance using machine learning techniques\cite{feng2017climate,nooteboom2018using,giffard2018fused,giffard20182018,giffard2018deep,li2018computing}. The second aim of this work is to propose a machine learning approach, using neural networks, to compute committor functions (section IV). The third result of this work is the demonstration of practicality of this approach on the example of a simple dynamics. We conclude by discussing the feasibility of the computation of a committor function using neural networks for the Jin and Timmerman model, and for more complex data sets related to other climate applications.

\section{The Jin and Timmermann model for ENSO}
The El-Ni\~{n}o phenomenon consists in an increase of the Sea Surface Temperature in the eastern equatorial Pacific Ocean and it is caused by a large-scale interaction between the equatorial Pacific Ocean and the global atmosphere. El-Ni\~{n}o is also related with the Southern-Oscillation phenomenon and the global phenomenon is called El-Ni\~no--Southern Oscillation (ENSO). In order to explain this phenomenon, in 1997 Jin introduced a simple dynamical model that accounts for the recharge-discharge mechanism which is at the basis of ENSO \cite{jin1997equatorial1,jin1997equatorial2}. This model was later extended by Timmermann \cite{timmermann2003nonlinear} and was related to the decadal amplitude changes of ENSO \cite{timmermann2002nonlinear}.

This model features the evolution of three variables:
\begin{enumerate}
\item $T_1$, the Sea Surface Temperature in the western equatorial Pacific Ocean,
\item $T_2$, the Sea Surface Temperature in the eastern equatorial Pacific Ocean,
\item $h_1$, the thermocline depth anomaly in the western Pacific. 
\end{enumerate}

The equations can be either deterministic or stochastic, a source of stochasticity being the variable wind stress related to the Walker circulation \cite{timmermann2002nonlinear,timmermann2003nonlinear}. After a change of variables\cite{roberts2016mixed}, from physical to dimensionless ones, the equations, introduced in \cite{timmermann2002nonlinear,timmermann2003nonlinear}, read\par
\begin{figure}
\centering
\includegraphics[scale=1.2]{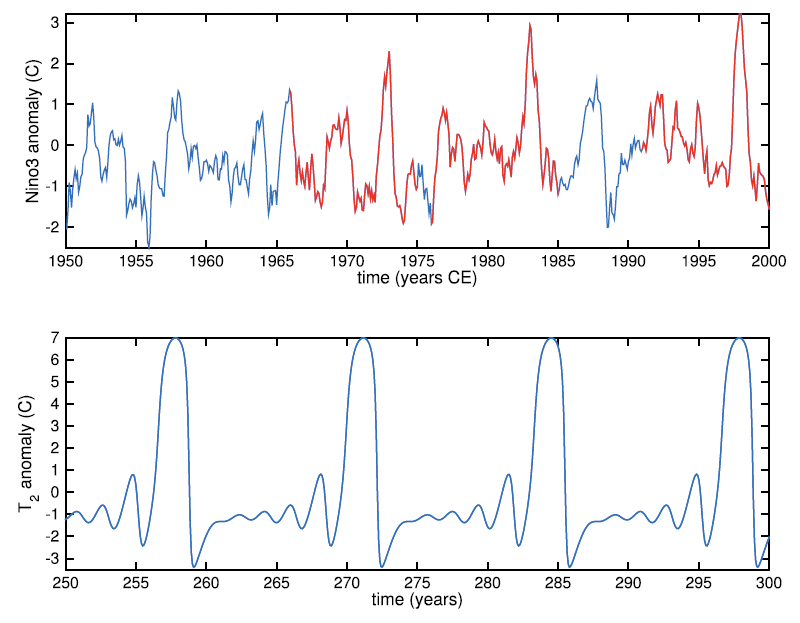}
\caption{Top plot: observed sea surface temperature anomalies, spatially averaged over the Ni\~{n}o-3 region. Bottom: eastern Pacific sea surface temperature anomalies simulated with the Jin and Timmermann model (figures from \cite{roberts2016mixed}).}
\label{ComparisonModelNinoIndex}
\end{figure}
\begin{figure}
\centering
\includegraphics[scale=0.15]{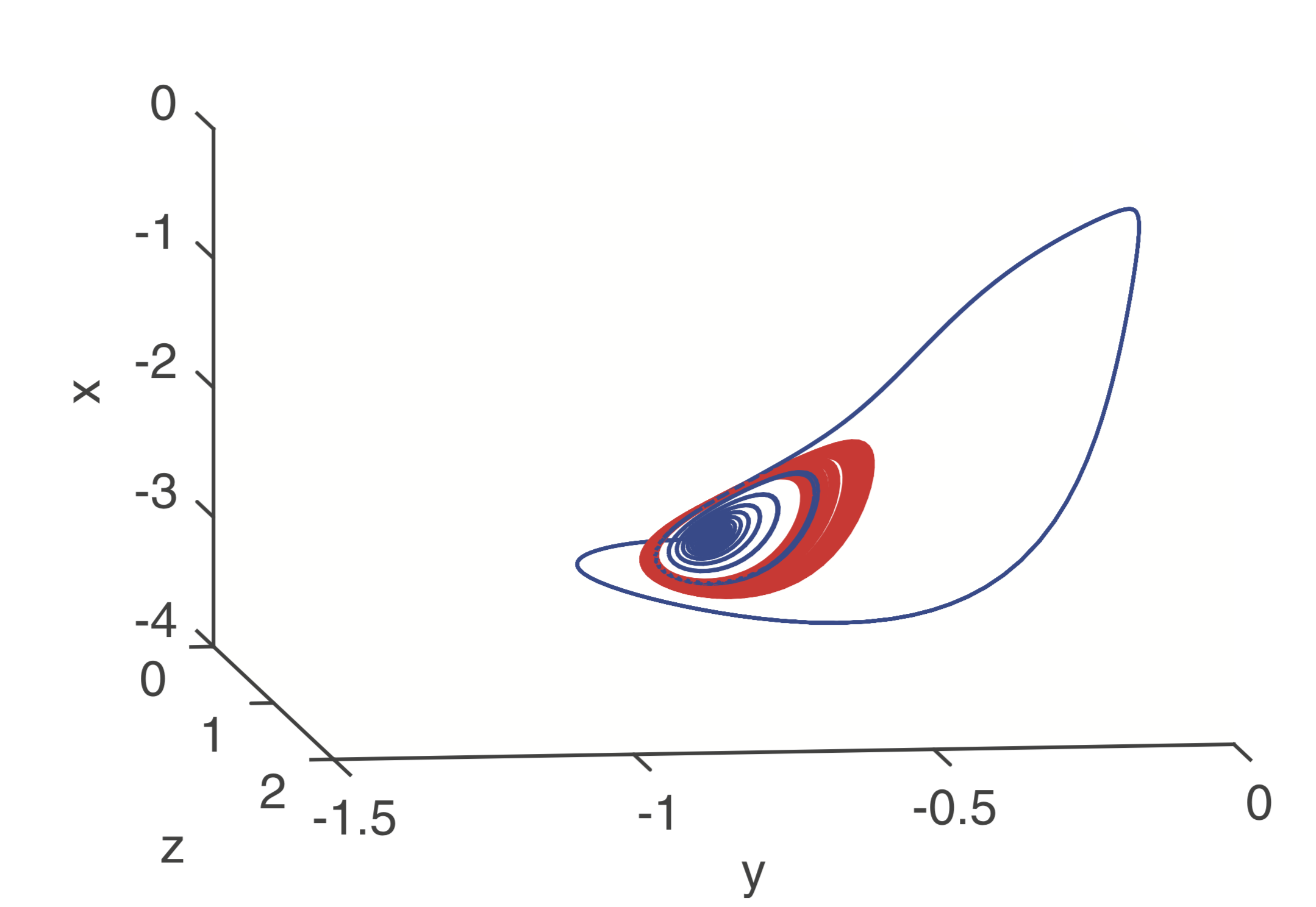}
\caption{The two intertwined attractors of the Jin and Timmermann model (periodic attractor in blue and chaotic one in red). Plot from \cite{guckenheimer2017predictability}.}
\label{TwoAttractors}
\end{figure}

\footnotesize
\begin{align}  
&\dot{x}=\rho\delta(x^2-ax)+x(x+y+c-c\tanh{(x+z)})-D_x(x,y,z) \xi_{t}, \nonumber \\    
&\dot{y}=-\rho\delta(x^2+ay)+D_y(x,y,z)\xi_{t}, \nonumber \\
&\dot{z}=\delta(k-z-\frac{x}{2}), 
\end{align}
\normalsize
where $x$ is $T_1-T_2$ divided by a reference temperature, $y$ is related to $T_1$, and $z$ is related to the thermocline depth $h_1$ (see \cite{roberts2016mixed}). The term $\xi(t)$ is a Gaussian white noise, \footnotesize $D_x(x,y,z)=[(1+\rho\delta)x^2+xy+cx(1-\tanh{(x+z)})]\sigma$ \normalsize and \footnotesize $D_y(x,y,z)=\rho\delta x^2\sigma$. \normalsize
The control parameters $[\delta,\rho,c,k,a,\sigma]$ are related to physical quantities \cite{roberts2016mixed}. We first describe the phenomenology when the noise is switched off ($\sigma=0$). For some parameter values, the system has only one attractor, a periodic orbit with oscillations that increase up to strong El-Ni\~{n}o events \cite{roberts2016mixed}. For other parameters the system has two different attractors: one periodic attractor that contains strong El-Ni\~{n}o events and one strange attractor without El-Ni\~{n}o events \cite{guckenheimer2017predictability}. These two attractors are intertwined with each other as illustrated in figure (\ref{TwoAttractors}). Figure (\ref{ComparisonModelNinoIndex}) shows a qualitative comparison of the eastern Pacific sea surface temperature anomalies for the periodic attractor with the El-Ni\~{n}o3 index. Both the measurements and the model display positive temperature anomaly excursions with a return time of approximately $20$ years. For this dynamics, we define a strong El-Ni\~{n}o event as any situation when $x$ becomes larger than the threshold $\epsilon=-1$.
Following \cite{guckenheimer2017predictability}, we have chosen these values for the parameters $[\delta,\rho,c,k,a]=[0.225423,0.3224,2.3952,0.4032,7.3939]$.

The level of stochasticity is controlled by the noise amplitude $\sigma$. For small $\sigma$, the dynamics can switch from one attractor to the other. The occurence of the next El-Ni\~{n}o event is then stochastic. Such a mode switching between attractors also occurs when the parameter $a$ is time periodic, mimicking a seasonal forcing \cite{guckenheimer2017predictability}. For large values of $\sigma$, the dynamics is completely dominated by the noise and the distinction between the two attractors becomes meaningless. 

\section{Committor Functions}\label{CommittorFunctions}
\begin{figure*}[htp]
\centering
\subfloat[][\emph{$\sigma=0$} \label{DeterministicX111}]
	{\includegraphics[width=.30\textwidth]{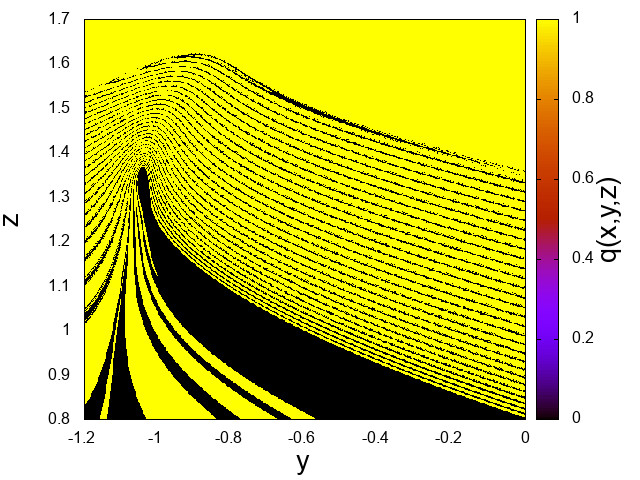} }\quad
\subfloat[][\emph{$\sigma=0.00005$}\label{X111Sigma00005}]
	{\includegraphics[width=.30\textwidth]{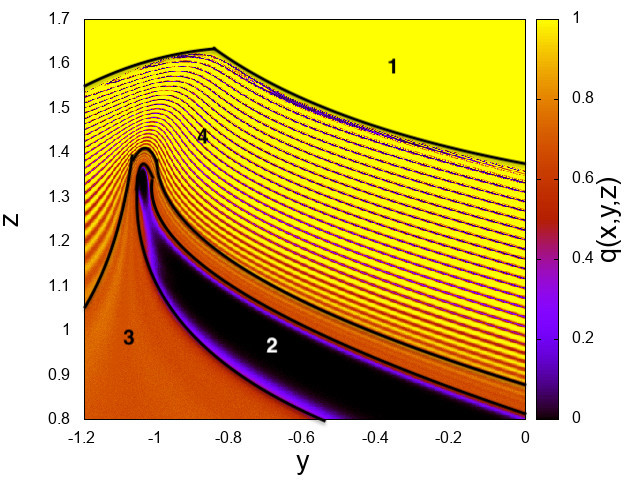}}\quad
\subfloat[][$\sigma=0.001$\label{X111Sigma001}]
	{\includegraphics[width=.30\textwidth]{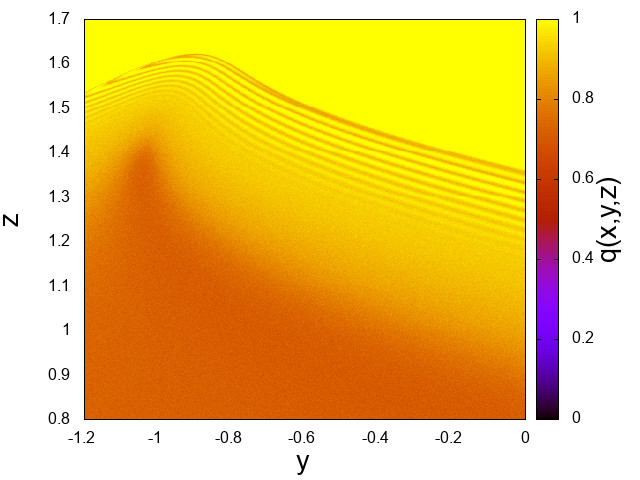} }
\caption{Colour plot of the committor $q$ versus $(y,z)$ for $x=-2.8310$, for $\sigma = 0$, $0.00005$ and $0.001$ respectively.}
\label{X111}
\end{figure*}
For a stochastic or a deterministic system, the probability $q(\mathbf{x})$ that a trajectory starting from the point $\mathbf{x}$ reaches a set $B$ of the phase space before another set $A$, as a function of the initial condition  $\mathbf{x}$  is called a committor function\cite{vanden2006transition, prinz2011efficient,weinan2005transition,thiede2019galerkin,schutte2015critical,schutte2011markov, lopes2019analysis}. It is thus a function on the phase space of the system. \\
Let us be more specific for the El-Ni\~{n}o prediction problem within the Jin and Timmerman model. $\mathbf{x} = (x,y,z)$ is the vector of the model phase space and $\left\{ \mathbf{X}(t)\right\}_{0\leq t \leq T}$ is one realisation of the dynamics. We consider the two sets \footnotesize $A=\{(\mathbf{x},t) \left| t \geq T \right. \}$ \normalsize and \footnotesize $B=\{(\mathbf{x},t)\left| x>\epsilon \right. \}$ \normalsize. 
We define the first hitting time of a set $C$ as\par
\footnotesize
\begin{equation}
\tau_C(\mathbf{x})=\inf\{t: \left( \mathbf{X}(t),t \right)\in C|\mathbf{X}(0)=\mathbf{x}\},
\end{equation}
\normalsize
The committor function $q(\mathbf{x})$ is the probability that the first hitting time of set $B$ is lower than the first hitting time of set $A$,  i.e.:\par
\footnotesize
\begin{equation}
q(\mathbf{x})=\mathbb{P}(\tau_B(\mathbf{x})<\tau_A(\mathbf{x})).
\label{eq:Committor}
\end{equation}
\normalsize
For the Jin and Timmerman model, the committor is thus the probability that the variable $x$ reaches the threshold $\epsilon$ before time $T$.



When the dynamics is a stochastic differential equation, which the case of the Jin and Timmerman model, one can prove that the committor function $q(\mathbf{x})$ is the solution of the Dirichlet problem \cite{weinan2005transition,thiede2019galerkin}\par
\footnotesize
\begin{equation}
\mathcal{L} q(\mathbf{x})=0 \rm{\ with\ } q(\mathbf{x})=0 \rm{\ if\ } \mathbf{x} \in A \rm{\ and\ } 
 q(\mathbf{x})=1\rm{\ if\ } \mathbf{x} \in B, 
\end{equation}
\normalsize
where $\mathcal{L}$ is the infinitesimal generator of the stochastic process ($\mathcal{L}$ is the adjoint of the Fokker-Planck operator)\par
\footnotesize
\begin{equation}
\mathcal{L}=\sum_i{a_i(\mathbf{x})\frac{\partial}{\partial x_i}(\cdot)}+\sum_{ij}{D_{ij}(\mathbf{x})\frac{\partial^2}{\partial x_i\partial x_j}}{(\cdot)}.
\end{equation}
\normalsize

Using a numerical simulation, 
one can simply generate $N$ different trajectories of length $T$ with initial condition $\mathbf{X}(0)=\mathbf{x}$. An estimate of the committor function $q(\mathbf{x})$ is simply $\frac{N_1}{N}$, where $N_1$ is the number of trajectories that reached the threshold $\epsilon$.

For the Jin and Timmermann model, we have chosen a value of $T$ slightly larger than the period of the periodic attractor. With this value we are at the predictability margin, with a value of $T$ of the order of the Lyapunov time, and of the order of the natural periodicity of El-Ni\~{n}o. This situation is thus analogous to trying to predict wether El-Ni\~{n}o will occur during next winter in the real climate dynamics. We also note that, since the averaged time required to switch from one attractor to the other one is greater than the period of periodic trajectories, each trajectory starting in one point of the periodic attractor almost certainly will reach the threshold $\epsilon=-1$. 

Figure (\ref{X111}) shows the committor function $q$, for different values of $\sigma$. As $q$ is a function of 3 variables  $(x,y,z)$, we have chosen to represent a cut of $q$ in the plane $x=-2.831$. Fig. (\ref{DeterministicX111}) shows $q$ for the deterministic dynamics ($\sigma=0$). For deterministic dynamics, as the future is completely determined by the initial condition, $q$ is equal to either $0$ or $1$. We see 3 regions. First, for larger values of $z$, in a large yellow area all trajectories reach the threshold and $q=1$. Second, in a thick black band, no trajectory reaches the threshold and $q=0$. Those two first regions are areas where the occurence or not of El-Ni\~{n}o is easily predicted. Third, everywhere else, we see very fine filaments of alternating yellow and black values. In this area, because of the sensitive dependance on the initial conditions, a small change of the initial conditions lead to a different outcome, and the occurence  of El-Ni\~{n}o is very difficult to predict.

When adding stochasticity, one clearly see by comparison of figures (\ref{DeterministicX111}), (\ref{X111Sigma00005}) and (\ref{X111Sigma001}) that the effect of a small noise blurs the visible structures of the deterministic case. For larger noise values, figure (\ref{X111Sigma001}) shows that while the deterministic predictability is lost for most initial points ($q \neq 0$ and $q \neq 1$), the committor function is smooth nearly everywhere. This means that the occurence of El-Ni\~{n}o is probabilistically predictable (the value of the probability can be determined in practise as it does not depend wildly on the initial condition).

The most interesting case is probably the one with the intermediate stochasticity value $\sigma = 0.00005$. In figure (\ref{X111Sigma00005}), we delineate 4 regions. First, two regions of perfect predictability, where the event will occur with probability $0$ (region 2) or $1$ (region 1), respectively. Second, regions 3) with good predictability properties where a value $0<q<1$ can clearly be predicted with very mild dependance with respect to initial conditions. We call this area the {\it probabilistically predictable region}. Third, regions 4) which are unpredictable in practice, because the strong dependance with respect to the initial condition prevent any practical prediction, either deterministic or probabilistic.  While regions 1), 2) and 4) are reminiscent of their deterministic counterparts, region 3) is not. It is a region where the stochasticity is large enough to smooth out the deterministic values of $q$. This occurs even at very low value of the stochasticity, probably in relations to regions leading to extremely unstable parts of the phase space, for instance for trajectories passing close to unstable fixed points or orbits. The existence of such features, and especially the new and most interesting {\it probabilistically predictable region}, region 3), should be generic for most prediction problems in climate dynamics.

\section{Learning the committor function with a neural network}

\begin{figure}
\centering
\includegraphics[scale=0.35]{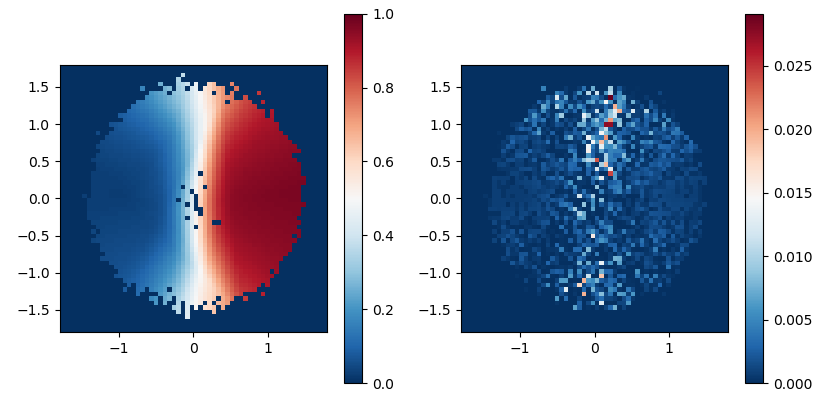}
\caption{Left: committor function for a two well gradient dynamics estimated using a neural network. Right: Error in the estimation of the committor function.}
\label{Committor_Machine_Learning}
\end{figure}
Estimating the committor function from its definition (\ref{eq:Committor}) 
requires a huge amount of data that is expected to increase exponentially with the phase space dimension. In order to cope with this issue, it could be very interesting to use committor regression, from data, using neural networks. In order to test this idea, we first note that the committor estimation amounts at regressing the parameter of a spatially dependent Bernoulli outcome between occurence $B$ with probability $q(\mathbf{x})$ and occurence $A$ with probability $1-q(\mathbf{x})$. As a consequence a natural loss function to be optimised by a neural network is the log-likelihood\par
{\scriptsize
\begin{equation}
C =  \frac{1}{N}\sum_{n=1}^N \left\{ y_n\log\left[1+\exp(-s(\mathbf{X_n}))\right] + (1-y_n)\log\left[1+\exp(s(\mathbf{X_n}))\right] \right\} \nonumber
\end{equation}}
\normalsize
\noindent where $q$ and $s$ are related through the logistic function \footnotesize $q(\mathbf{x}) = 1/\left[1+\exp(-s(\mathbf{x}))\right]$\normalsize. The data $\left\{ \right(\mathbf{X}_n,y_n\left) \right\}_{1 \leq n \leq N}$ couples each represent a phase space point $\mathbf{X}_n$ corresponding to an initial condition of the dynamics and a value $y_n$ equal to either $1$ if the trajectory reaches $B$ before $A$, or $0$ otherwise. The function $s$ is determined as the minimiser of $C$. The committor function $q$ is then computed from the relation between $q$ and $s$.

We have tested this approach on a simple gradient stochastic dynamics with two degrees of freedom. The deterministic part of the dynamics has two point attractors. In a limit of small noise, we look for the committor function defined as the probability that the trajectory reaches a neighbourhood of the first attractor before reaching the neighbourhood of the second one. For this simple example, we trained the neural network with simulated trajectories. The neural network model had a standard 3-layer fully connected Multilayer Perceptron (MLP) architecture (32, 64 and
1 neurons respectively) and Rectified Linear Unit (ReLU) activation functions for the hidden layers. 
Figure (\ref{Committor_Machine_Learning}) shows the estimated committor and the computed error, using as a benchmark a committor computed from its definition (\ref{eq:Committor}) and using an extremely long data set. This figure clearly demonstrates the ability of the neural network to learn precisely the committor for this simple example.

In future works, we will use this machine learning approach to learn the committor function for El-Ni\~{n}o, within the Jin and Timmerman model. Our main aim will be to demonstrate the efficiency of this approach and to estimate the required amount of data for genuine climate applications.

\newpage

\section*{Acknowledgments}
This work has received funding through the ACADEMICS grant of the IDEXLYON, project of the Université de Lyon, PIA operated by ANR-16-IDEX-0005.
The computation of this work were partially performed on the PSMN platform of ENS de Lyon. During this project, we benefitted from scientific discussions with Patrice Abry, Pierre Borgnat and Charles Edouard Brehier.

\bibliographystyle{ieeetr}
\bibliography{Bibliography}

\end{document}